\documentclass[sigconf, authorversion,nonacm]{acmart}
\usepackage{hyperref}
\usepackage{graphicx}
\usepackage{algorithmic}
\usepackage{algorithm}
\usepackage{xcolor}

\copyrightyear{2023} 
\acmYear{2023} 
\setcopyright{rightsretained} 
\acmConference[SC-W 2023]{Workshops of The International Conference on High Performance Computing, Network, Storage, and Analysis}{November 12--17, 2023}{Denver, CO, USA}
\acmBooktitle{Workshops of The International Conference on High Performance Computing, Network, Storage, and Analysis (SC-W 2023), November 12--17, 2023, Denver, CO, USA}
\acmDOI{10.1145/3624062.3624232}
\acmISBN{979-8-4007-0785-8/23/11}


\begin{document}
\title{Challenges and Opportunities in the Co-design of Convolutions and RISC-V Vector Processors}
%
%
\author{Sonia Rani Gupta} 
\affiliation{
\institution{Chalmers University of Technology} 
\city{Gothenburg} 
\country{Sweden}
}
\email{soniar@chalmers.se}

\author{Nikela Papadopoulou}
\affiliation{
\institution{Chalmers University of Technology} 
\city{Gothenburg} 
\country{Sweden}
}
\email{nikela@chalmers.se}

\author{Miquel Pericàs}
\affiliation{
\institution{Chalmers University of Technology} 
\city{Gothenburg} 
\country{Sweden}
}   
\email{miquelp@chalmers.se}

%

%
\begin{abstract}
The RISC-V "V" extension introduces vector processing to the RISC-V architecture. Unlike most SIMD extensions, it supports long vectors which can result in significant improvement of multiple applications. In this paper, we present our ongoing research to implement and optimize a vectorized Winograd algorithm used in convolutional layers on RISC-V Vector(RISC-VV) processors. Our study identifies effective techniques for optimizing the kernels of Winograd on RISC-VV using intrinsic instructions, and showcases how certain instructions offer better performance. Our co-design findings suggest that the Winograd algorithm benefits from vector lengths up to 2048 bits and cache sizes up to 64MB.

We use our experience with Winograd to highlight potential enhancements for the standard that would simplify code generation and aid low-level programming. Finally, we share our experience from experimenting with forks of gem5 for RISC-VV and stress the importance of a mature software ecosystem, to facilitate design space exploration and architectural optimization.

Our study identifies effective techniques for optimizing the kernels of Winograd on RISC-VV using the available intrinsic instructions and showcases that certain instructions offer better performance to the vectorized algorithm.
Furthermore, our co-design study reveals that the Winograd algorithm benefits from vector lengths up to 2048 bits and cache sizes up to 64MB.



\end{abstract}


\maketitle          

\section{Introduction}

RISC-V, an open standard Instruction Set Architecture (ISA), has gained significant interest in the High-Performance Computing (HPC) and Artificial Intelligence (AI) communities as an alternative to proprietary architectures, such as x86, ARM, and CUDA. One of the main advantages of RISC-V is its customizable nature, allowing for bespoke designs to meet specific HPC/AI needs. The "V" extension of RISC-V adds vector processing, allowing implementations to process large amounts of data in a parallel and efficient manner. 

Vector architectures have evolved considerably since they were first introduced in early systems such as the Cray-1~\cite{cray1}. 
Modern vector extensions can broadly be classified into two groups. In the first group, fixed-length vector extensions are common for media processing such as audio or video processing. This includes vector extensions such as Intel x86 MMX, SSE, AVX, AVX2, and AVX512~\cite{Intel}, and ARM NEON~\cite{Arm-neon}, in which vector registers are of fixed length. On the other hand, ARM SVE~\cite{ARMSVE} and RISC-V Vector Extension (RISC-VV)~\cite{RISC-VSpec} are extensions to the respective ISAs based on long vectors and vector length agnostic ISAs, and can result in significant performance improvements for a wide range of applications, including scientific computing, machine learning, and multimedia processing. One of the key features of ARM-SVE and RISC-VV is their support for long vector lengths, which enables high parallelism with lower complexity for the architecture. 
This feature also improves the energy efficiency by reducing the number of instructions required to complete a task, thereby reducing the energy consumed by the processor's front end, which is a significant concern for servers with power caps and mobile devices with limited battery life. Another key feature is the vector length agnostic (VLA) nature of these ISAs, which provides code portability across implementations with different hardware vector lengths. Vector processors based on VLA are featured in various designs such as EPI's Vector accelerator~\cite{EPILink}, Fujitsu A64FX~\cite{Fugaku} and ARM Neoverse-V1~\cite{noverse-1}.


The RISC-V Vector extension has been implemented with vector lengths up to 16384 bits~\cite{cavalcante2019ara,minervini2023vitruvius+}. The supported vector length is an important feature to improve the efficiency of heavily compute-intensive applications, or applications with regular, high data-level parallelism \cite{ramirez2020risc}. Several works~\cite{gomez2023hpcg,gomez2021efficiently} focus on porting and optimizing HPC kernels and applications on long-vector architectures. In our previous work \cite{gupta2022accelerating}, we have focused on convolutional neural networks (CNNs) on long vector architectures. We have ported and optimized the im2col+GEMM and Winograd algorithms, found in the compute-intensive convolutional layers, concluding that longer vector lengths (up to 8192 bits) are beneficial to the performance of CNNs. \looseness=-1

In this work, we extend our previous work\cite{gupta2022accelerating} with an optimized implementation of the Winograd~\cite{ala2022winograd,winogradcoppersmith} algorithm using RISC-VV, and use our optimized implementations of the convolutional layers, with both im2col+GEMM, and Winograd, for a co-design study on a simulated RISC-V processor featuring the "V" extension. We focus on the RISC-V Vector Extension v1.0, and use the intrinsics available in the RISC-V LLVM/Clang toolchain \cite{EPI-builtin} from the European Processor Initiative (EPI) \cite{EPILink}, to vectorize the kernels in Winograd. 


In summary, this paper makes the following contributions:
\begin{itemize}
   \item We present our experience in vectorizing and optimizing the Winograd kernels using RISC-VV in a vector length agnostic way. We showcase effective techniques to vectorize the Winograd kernels, leveraging the available intrinsic instructions. We demonstrate that workarounds to avoid indexed vector load/stores can offer significant performance 
   \item  We perform a co-design study for the VGG16 network model, using our vectorized and optimized Winograd algorithm for the convolutional layers. Considering the vector length and L2 cache size in our exploration, we show that Winograd is able to utilize vector lengths up to 2048 bits, improving performance by $\sim$1.4$\times$ compared to a vector length of 512 bits, however longer vector lengths do not offer further gains. Additionally, Winograd scales up to 64MB of L2 cache size, but does not require larger cache sizes. 
    Through a similar co-design study for the YOLOv3 network model, which uses both im2col+GEMM and Winograd to implement the convolutional layers, we show that vector lengths of 4096 bits improve performnace by 1.76$\times$ and a L2 cache size of 256MB improves performance further by 1.5$\times$ (i.e. 2.6$\times$ in total), compared to a 512-bit long vector architecture with 1MB of L2 cache size. 
  \item Finally, we use the Winograd algorithm as a case study to compare the RISC-VV and ARM-SVE ISAs. 
   We observe similar performance and performance trends, on both RISC-VV and ARM-SVE architectures. We also discuss our experience with simulators/emulators available for RISC-VV and highlight the necessity of mature toolchains and alignment between the compilers and tools. 

\end{itemize}

\label{sec:intro}

    
\section{Implementation of Convolutions}
The convolutional layer can be implemented using different algorithmic implementations such as Direct, im2col+GEMM, Winograd, and FFT. The Direct convolution implementation is mainly used for 1x1 kernel size, as its memory footprint increases with larger kernel sizes and it exhibits low data reuse, although recent approaches have optimized the Direct convolution for SIMD \cite{zhang2018high} and long-vector architectures \cite{santana2023efficient}. The Winograd implementation reduces the computational complexity but is mainly helpful in 3$\times$3 or 5$\times$5 kernel sizes, because of a numerical inaccuracy issue for large kernel sizes. FFT also has the advantage of reducing the complexity but is beneficial with bigger kernel sizes. On the other hand, im2col+GEMM can be generically applied and is efficient on most modern architectures \cite{surveycnnimplementation}. The latest, state-of-the-art convolutional network  models mainly employ kernel sizes of 1$\times$1, 3$\times$3, or 5$\times$5, making Winograd and im2col+GEMM better choices for implementing the convolutional layer. 

In this work, we port and optimize the Winograd algorithm from the NNPACK \cite{nnpack} software library, to implement convolutional layers with 3$\times$3 kernel size, of stride equal to 1. In our evaluation, for the remaining convolutional layers,  with different kernel sizes or strides, we use the im2col+GEMM implementation from the Darknet framework~\cite{darknet13}. 

The Winograd algorithm involves transforming both the input data and filter data, performing tuple multiplication on the resulting, transformed matrices, and then transforming the output back to its original form. The Winograd algorithm, as implemented in NNPACK, uses an 8$\times$8 input tile and a 3$\times$3 filter to produce a 6$\times$6 output. Vectorization with longer vector lengths requires employing a scheme of inter-tile parallelization, across the input/output channels by taking an 8$\times$8 tile from each channel, as explained in our previous work~\cite{gupta2022accelerating}. In this work, we use the same inter-tile parallelization scheme as a base for vectorizing the Winograd algorithm in a VLA way, to utilize the longer vector lengths available on the RISC-VV architecture. 

The im2col+GEMM algorithm relies on the im2col kernel to transform the input into column matrices, and on the common GEMM kernel to execute matrix multiplication on the input matrix and the transformed column matrix. We are vectorizing and optimizing im2col and GEMM kernels in a VLA way to utilize the longer vector lengths efficiently as shown in~\cite{gupta2022accelerating}. 
We are using our optimized implementation for these kernels for the convolutional layers, where we can not use the Winograd implementation.

\label{background}

\section{Winograd Implementation on RISC-VV}


As discussed in Section~\ref{background}, we have used an inter-tile parallelism approach across the input/output channels, to be able to vectorize the Winograd algorithm for longer vector lengths. To utilize 512-bit vector registers, we use 8$\times8$ tiles from 4 channels. In order to fully utilize longer vector lengths, e.g. 8192-bit vector lengths, we need to use equivalently more channels, e.g. to use 64 channels. In the convolutional layers where the number of channels is more than 4, we enable inter-tile parallelism for RISC-VV. Further, to vectorize the tuple multiplication, which can employ up to 8192-bit-long vectors, we now increase the number of blocks to 64, with 4 elements in each block. Thus, we can vectorize and use the Winograd algorithm with long vectors with RISC-VV. To manually vectorize all the kernels of the Winograd algorithm, we use the intrinsics available in the EPI LLVM/Clang toolchain for the RISC-V Vector Extension. We note that we also use the compiler auto-vectorization capabilities to let all the kernels to auto-vectorize. After examining the produced assembly code, we have observed that only one loop out of all kernels is auto-vectorized. Therefore, for further experimentation, we use manually vectorized kernels.

During the process of manually vectorizing with intrinsic instructions, as well as in our evaluation (see Section~\ref{results}), we demonstrate the use of effective intrinsics (and RISC-VV instructions) for the vectorization of the Winograd algorithm, which we have evaluated with small code snippets. Based on this experience, we hint towards best-performing instructions, as well as towards some enhancements in RISC-VV, which would boost low-level/assembly-level programmability. In the next paragraphs, we highlight the most important parts of our optimization process.


\paragraph{Input transformation.}
In the input transformation kernels of Winograd, approximately 30 instructions which transform the input data and produce the transformed output are used at
    6 places, with different input data sets.  These instructions produce the intermediate output before producing the final, transformed output. In this scenario, it would be more practical to incorporate these instructions into a function and invoke it whenever transformation is necessary. However, this function must update output registers with the transformed output, thus requiring that output registers be passed as a reference. As RISC-VV does not support pointers with vector datatypes, while optimizing the input transformations using manual vectorization on RISC-VV, we are forced to incorporate these 30 instructions in the main function at 6 different places, making the code lengthy, error-prone, and less efficient. The same holds for the kernel and output transformation. Although using macros as an alternative to functions can help improve the readability of the code, the extended need for intermediate registers can still cause register spilling. Although the compiler can optimize register usage through heuristics, the programmer may not be able to avoid register spilling. 
    As a result, we conclude that having pointers with vector data types in the "V" extension will improve the programmability and will also help reduce the chances of register spilling.
    

\begin{algorithm}[]
\caption{Tuple multiplication code snippet from Winograd showcasing the indexed load work-around in RISC-VV}

\begin{algorithmic}[1]

\STATE $ind\gets 0$ 
\STATE  VL = get vector length
\STATE $elements = 4$
\STATE $rem = VL/elements$
\STATE $num\gets blocks*4$ //64 blocks
\STATE $iteration\gets num/64$ 
\FOR{$i\gets 0$, $i<rem$, $i+=1$}
\FOR{$j\gets 0$, $j<4$, $j+=1$}
\STATE $index[ind] = j$
\STATE $ind++$;
\ENDFOR
\ENDFOR

\FOR{$itr\gets 0$, $itr<num$ }
        \STATE $gvl = setvl(num-itr)$
        \STATE $index\_vec\gets index[0:gvl]$
        \FOR{$k\gets0$, $k<iteration$, $k+=1$}
        \FOR{$j\gets 0$, $j< inputchannels$, $j+=1$}
            \STATE $b0\_vec \gets B[itr+(j*VL)]$ //load filter matrix
            \FOR{$i\gets 0$, $i<64$, $i+=4$}
            \STATE $a0\_vec \gets indexed\_load(A[(i+(64*k))+(j*VL)],index\_vec, gvl])$
            \STATE $acc\_vec[i] \gets vfmacc(acc\_vec[i], a0\_vec, b0\_vec, gvl)$
        \ENDFOR
        \ENDFOR
        \STATE //Store results in the resultant matrix in a contiguous way
    \ENDFOR
    \STATE $itr+=gvl$;
\ENDFOR
\end{algorithmic}
\label{algo:winogradRISCVV}
\end{algorithm}

\begin{algorithm}[]
\caption{Tuple multiplication code snippet from Winograd using the slideup instructions in RISC-VV}
\begin{algorithmic}[1]

\STATE $ind\gets 0$ 
\STATE  VL = get vector length
\STATE $iteration\gets num/64$ 
\FOR{$itr\gets 0$, $itr<num$ }
        \STATE $gvl = setvl(num-itr)$
        \STATE $index\_vec\gets index[0:gvl]$
        \FOR{$k\gets0$, $k<iteration$, $k+=1$}
        \FOR{$j\gets 0$, $j< inputchannels$, $j+=1$}
            \STATE $b0\_vec \gets B[itr+(j*VL)]$ //load filter matrix
            \FOR{$i\gets 0$, $i<64$, $i+=4$}
            \STATE $a0\_vec \gets load(A[(i+(64*k))+(j*VL)], gvl])$
            \FOR{$ind\gets0$, $4*ind<=gvl/2$, $ind+=1$}
                \STATE $a0\_vec \gets Slideup(a0\_vec, 4*ind, gvl)$
            \ENDFOR
            \STATE $acc\_vec[i] \gets vfmacc(acc\_vec[i], a0\_vec, b0\_vec, gvl)$
        \ENDFOR
        \ENDFOR
        \STATE //Store results in the resultant matrix in a contiguous way
    \ENDFOR
    \STATE itr+=gvl;
\ENDFOR
\end{algorithmic}
\label{algo:winogradRISCVVSlideup}
\end{algorithm}

\begin{algorithm}[]
\caption{Transpose code snippet from Winograd for RISC-VV}
\begin{algorithmic}[1]
\STATE $VL$ = vector length
\STATE $gvl$ = granted vector length
\STATE $V0, V1, V2, V3$ //Vector registers
\STATE $vin0123[0]\gets V0$
\STATE $vin0123[VL]\gets V1$
\STATE $vin0123[2*VL]\gets V2$
\STATE $vin0123[3*VL]\gets V3$
\STATE $ind\gets 0$ 
\FOR{$j = 0$, $j < 4$, $j+=1$}
\STATE $index[ind] = ((j * VL))$
\STATE $ind++$
\ENDFOR
\STATE  //load index in $index\_vec$
\STATE $trans\_vec[0:gvl]\gets$  $indexed\_load(\&vin0123[0], index\_vec, gvl)$
\end{algorithmic}
\label{algo:transposeriscvv}
\end{algorithm}

\begin{algorithm}[]
\caption{Transpose code snippet using strided from Winograd for RISC-VV}
\begin{algorithmic}[1]
\STATE $VL$ = vector length
\STATE $gvl$ = granted vector length
\STATE $V0, V1, V2, V3$ //Vector registers
\STATE $vin0123[0]\gets strided\_store(V0, stride,gvl)$
\STATE $vin0123[1]\gets strided\_store(V1,stride, gvl)$
\STATE $vin0123[2]\gets strided\_store(V2, stride, gvl)$
\STATE $vin0123[3]\gets strided\_store(V3,stride,gvl)$

\STATE $trans\_vec[0:gvl]\gets load(\&vin0123[0], gvl)$
\end{algorithmic}
\label{algo:transposeriscvvstride}
\end{algorithm}

\begin{figure}[t]
  \centering
\includegraphics[width=\linewidth]{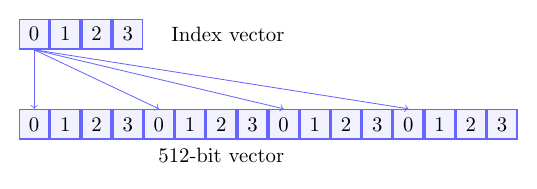}
  \caption{
  Indexed load for packing quadword in a 512-bit vector register}
  \label{fig:tuplemult}

 \end{figure}

\paragraph{Tuple multiplication.} In the tuple multiplication kernel from the Winograd algorithm, the logic requires reading a block of 4 $\times$ 32-bit elements (quadword) from the transformed input matrix, to perform the element-wise multiplication with the loaded transformed filter matrix. 
For this task, which is illustrated in Figure~\ref{fig:tuplemult}, there is no matching intrinsic available, thus we have explored different intrinsic instructions to fill in the vector register from the quadword. 
    In our first attempt, we have used the indexed vector load intrinsic to fill the vector register, which we present in Algorithm~\ref{algo:winogradRISCVV}. 
    In line 18, we employ an indexed vector load operation to retrieve a quadword from the $A$ matrix. To use the indexed vector load, we generate indices in lines 5 to 10 and load them into the index vector $index\_vec$. We then utilize the index vector as the second argument when loading data from the $A$ matrix to the $a0\_vec$ vector register. Our preliminary evaluation showed that these indexed load/store (gather/scatter) instructions are expensive, therefore, we have taken an alternative approach, using the slideup intrinsic instruction, which we illustrate in Algorithm~\ref{algo:winogradRISCVVSlideup}. 
    In line 10, we use a vector load operation to retrieve data from the $A$ matrix in the $a0\_vec$ vector register. Then we use the slideup intrinsic instructions from line 12 to line 14 to fill the vector with the quadword data. We have compared tuple multiplication with the two approaches, for a 100 iterations, finding that tuple multiplication using the slideup intrinsic is $\sim$2.3$\times$ faster than using the indexed load/store instructions (see Section~\ref{sec:setup} for the experimental setup).

\begin{figure}[!t]
  \centering
  \includegraphics[width=\linewidth]{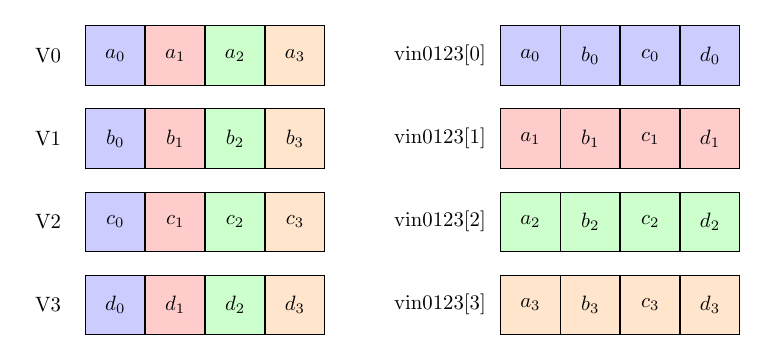}
 
  \caption{
  Transpose the elements of the vectors residing in vector registers V0, V1, V2, and V3.}
  \label{fig:tuplevec}

 \end{figure}

\paragraph{Transpose operation.} For the transformations in the Winograd algorithm, we perform the transpose operation on an array of 4 vectors, as shown in Figure~\ref{fig:tuplevec}.  However, in  RISC-VV, currently, no specific instructions are available to perform these operations (transposing the two vectors is available as EPI custom extensions~\cite{EPI-builtin}, but not in the standard "V" extension). We, therefore, implement a solution that uses temporary buffers and additional store and gather-load instructions as shown in Algorithm~\ref{algo:transposeriscvv}. To transpose 4 vectors, the process involves storing the vector data into buffer $vin0123$ using memory operations in lines 2 to 5. Subsequently, we generate indices and load them into the $index\_vec$ vector register in lines 8 to 12. Finally, we use an indexed load vector instruction in line 13 to load the transposed values from the buffer into a vector register. \looseness=-1 

As an alternative approach, we have also used strided stores to store the intermediate  vector results in the buffer $vin0123$, using a stride of 16  (4 single-precision floating-point elements) to transpose the vectors. We demonstrate this approach in lines 4 to 7 of Algorithm~\ref{algo:transposeriscvvstride}. In this approach, the already transposed data is loaded in line 8 using a contiguous load intrinsic instruction. Comparing the performance of the two approaches for transformations (see Section~\ref{sec:setup} for the experimental setup), we did not observe any significant performance difference. 
    We note, however, that both cases require memory operations, i.e. scatter/store and gather/load instructions, which put a burden on performance, as well as programmability, forcing the developer to implement more complex solutions. Therefore, we advocate for an extension of the RISC-VV with vector transpose instructions, that would eliminate the need for memory operations. 
    

    
    
\section{Experimental Setup}
\label{sec:setup}

For our work, we use the EPI-Builtins~\cite{EPI-builtin} to vectorize the kernels of the Winograd algorithm from the NNPACK~\cite{nnpack} package with VLA vectorization on RISC-VV. We use the EPI fork of the LLVM~\cite{llvm-epi} Clang cross-compiler version 17.0.0 for RISC-VV with -O3 optimization flag to compile the algorithm and the network models. To validate our algorithms during development, we use Spike \cite{SPIKE}, a RISC-V ISA simulator that supports vector lengths up to 4096 bits and supports the "V" extension v1.0. To evaluate the performance of our optimized algorithm and explore the impact of hardware parameters, such as the vector length and the L2 cache size, on performance, we use a fork of the gem5~\cite{plct-gem5} simulator, a cycle-accurate simulator configured with the RISC-V in-order RiscvMinorCPU CPU model, with a core frequency of 2GHz, for RISC-VV, targeting the RISC-VV 1.0. The memory subsystem of RiscvMinorCPU is configured with two levels of data cache, 64kB of L1 and 1MB of L2 cache. Note that this fork of gem5 models a constant latency for all the vector instructions. In practice, the latency of the instructions will vary with the implementation of RISCV-V. 

We evaluate our optimized Winograd algorithm in the context of the convolutional layers found in two network models, YOLOv3~\cite{yolov3}, an object detection network model, and VGG16, an image classification network model, from the Darknet \cite{darknet13} framework, on an inference task with a 768$\times$576 pixels input image.

\section{Results}
The convolutional layers of YOLOv3 use both $1\times1$ and $3\times3$ kernel sizes, with stride 1 and higher,  therefore we use our optimized Winograd algorithm to implement the convolutional layers with kernel sizes $3\times3$ and stride equal to 1, and use our optimized im2col+GEMM algorithm \cite{gupta2022accelerating} for the remaining layers, in what we call the \emph{hybrid approach}. We then compare the performance achieved by the hybrid approach compared to the pure im2col+GEMM approach, where all layers are implemented with im2col+GEMM. To avoid extreme simulation times, and without loss of generality, we simulate only the first 20 layers of the network model, out of which 15 are convolutional layers. Compared to the pure im2col+GEMM approach, the hybrid approach achieves an $\sim8\%$performance improvement on a simulated RISC-V processor with a vector length of 2048 bits and an L2 cache of 1MB, using gem5. We conclude that the Winograd algorithm runs faster compared to im2col+GEMM on RISC-V. The reason for the limited performance improvement is that, out of the 20 layers, only 5 layers use the Winograd algorithm and contribute to the performance improvement, as 3 layers have a stride equal to 2, 6 layers have a $1\times1$ kernel size, the first layer uses only 3 input channels, and 5 layers are not convolutional layers. Finally, for performance validation, we compare the performance achieved on RISC-VV to the performance we have previously achieved with ARM-SVE on gem5, with the same CPU model, finding that Winograd performs the same on both vector architectures. 

\begin{figure}[!t]
  \centering
  \includegraphics[width=\linewidth]{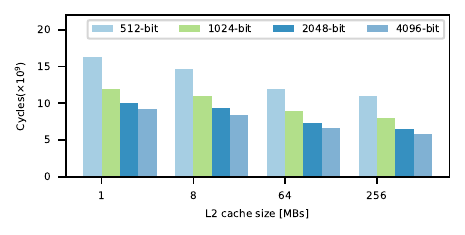}
  \caption{
  Impact of vector lengths and L2 cache size with Winograd on RISC-VV@gem5 for YOLOv3.}
  \label{fig:YOLOV3longercachewinograd}
  \vspace{-1.3em}
\end{figure}

We then perform a co-design study to assess the impact on performance of tuning the hardware parameters of RISC-VV using the gem5 simulator. The available fork of the gem5 simulator for RISC-VV allows us to simulate only up to 4096 bits long vector lengths. We simulate vector lengths of 512 bits up to 4096 bits, and L2 cache sizes of 1 up to 256 MB. We demonstrate our results in Figure~\ref{fig:YOLOV3longercachewinograd}. 
We observe a speedup of $\sim$1.76$\times$ by increasing the vector length from 512 bits to 4096 bits. Along with the indexed load, the L2 cache miss rate is another factor that limits the scalability with longer vector lengths. As shown in Table~\ref{tab:L2miss}, increasing the vector length makes the L2 cache miss rate increase, as longer vectors process more data at once and thus require bigger caches to pump more data to the vector units for processing. By increasing the cache size from 1MB to 256MB for different vector lengths, we observe a speedup of 1.5$\times$ from 1MB to 256MB of L2 cache size for 512-bit and 1024-bit vector lengths, and 1.54$\times$ and 1.6$\times$ for 2048-bit and 4096-bit vector lengths respectively.

 \begin{table}[tb] 
\centering
\caption{L2 cache miss rate of YOLOv3 inference on RISC-VV for 1MB of L2 cache.}
\begin{tabular}{ccc} 
\hline 
\textbf{Vector length}  & \textbf{L2 cache miss rate(\%)} \\ \hline
512-bit      &39              \\ 
1024-bit     &47          \\
2048-bit   & 50\\
4096-bit    & 52\\\hline

\end{tabular}
\label{tab:L2miss}
\end{table}

\begin{figure}[!t]
  \centering
  \includegraphics[width=\linewidth]{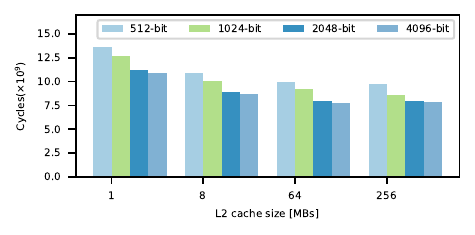}
  \caption{
  Impact of vector lengths and L2 cache size with Winograd on RISC-VV@gem5 for VGG16.}
  \label{fig:VGG16longercachewinograd}
  \vspace{-1.3em}
\end{figure}

We additionally evaluate our implementation of Winograd with the VGG16 network model, where all the layers are convolutional with a kernel of size $3\times3$ and a stride of 1, we thus use Winograd for all the layers of the network model. Comparing the performance of VGG16 using Winograd against VGG16 using im2col+GEMM for all layers, on our simulated RISC-VV architecture with gem5, we observe a speedup of $1.2\times$ for a vector length of 2048 bits and 1MB of L2 cache. We then perform a similar exploration of hardware parameters as for YOLOv3. We showcase the results in Figure~\ref{fig:VGG16longercachewinograd}. We observe that as we increase the vector length from 512 bits to 4096 bits, the performance improves by 1.4$\times$. However, we do not see a significant improvement beyond 2048 bits. For further investigation, we examine the impact of the L2 cache miss rate for different vector lengths, which we present in Table~\ref{tab:L2missVGG16}. We observe that the L2 cache miss rate is unaffected by the vector length and does not limit performance scaling beyond 2048 bits. Additionally, from analyzing the gem5 statistics, we observe that there is no significant difference in the number of instructions from 2048-bit to 4096-bit vector lengths, which indicates that our Winograd algorithm does not utilize very long vector lengths. Due to reduced computational complexity, Winograd reduces the required floating point operations for implementing the convolutional layer and therefore can provide performance benefits with moderately large vector lengths. Also, our roofline analysis (see Section~\ref{Roofline} shows that the Winograd algorithm is memory-bound on the simulated architecture, which limits its scalability with longer vector lengths. 
To observe the impact of bigger cache sizes, we increase the cache sizes from 1MB to 256MB for different vector lengths and observe performance improvement of $\sim$1.3$\times$ from 1MB to 64MB for different vector lengths. Beyond 64MB of cache, no significant performance improvement is observed. This concludes that our Winograd implementation does not require very large L2 caches and can scale with moderately large cache sizes. This observation is in accordance with our conclusions from optimizing Winograd with ARM-SVE.  

 \begin{table}[tb] 
\centering
\caption{L2 cache miss rate of VGG16 inference on RISC-VV for 1MB of L2 cache.}
\begin{tabular}{ccc} 
\hline 
\textbf{Vector length}  & \textbf{L2 cache miss rate(\%)} \\ \hline
512-bit      &80              \\ 
1024-bit     &84          \\
2048-bit   & 85\\
4096-bit    & 82\\\hline

\end{tabular}
\label{tab:L2missVGG16}
\end{table}




\label{results}

\section{Roofline Analysis}
\begin{figure}[!t]
  \centering
  \includegraphics[width=\linewidth]{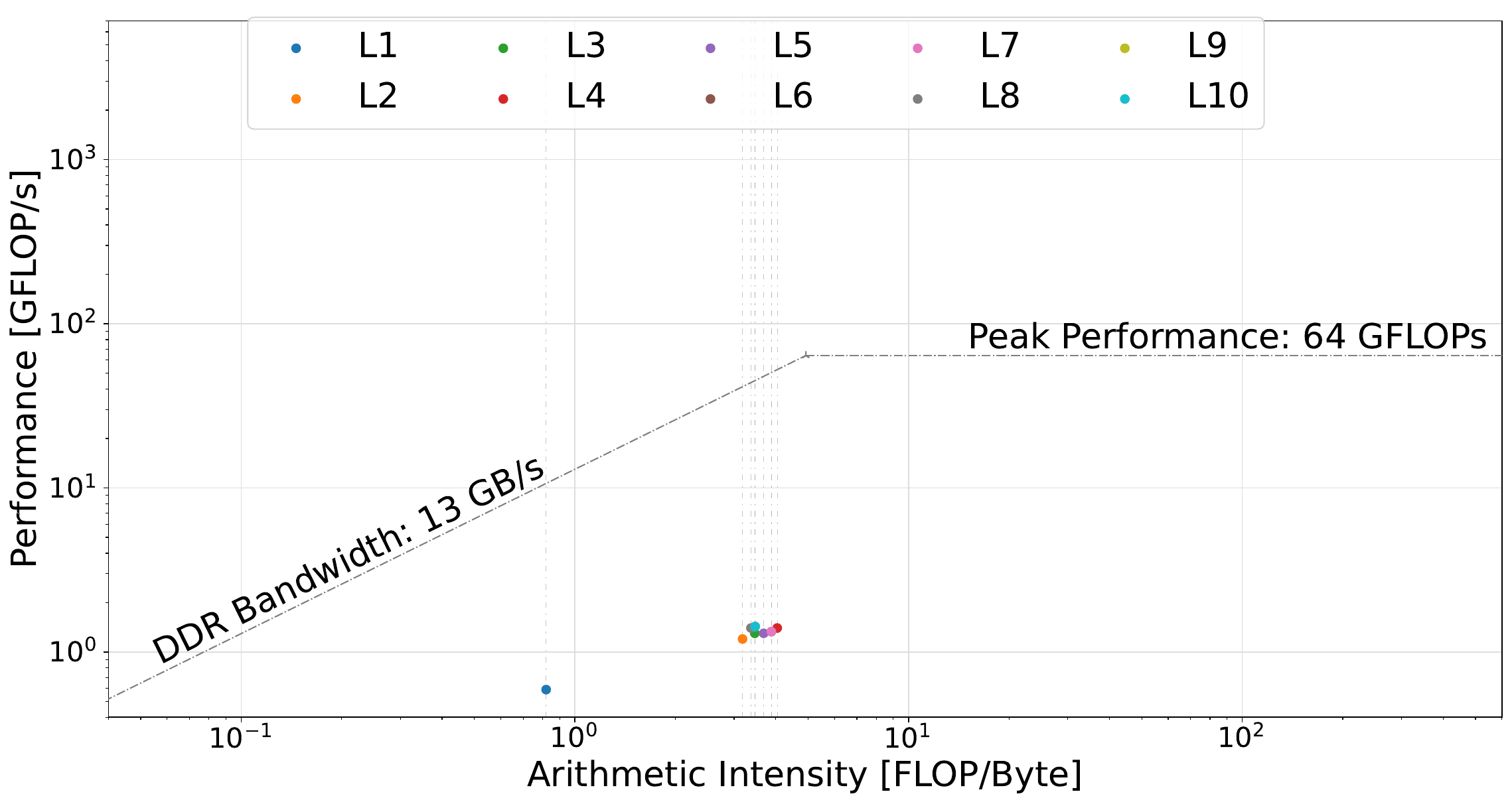}
  \caption{
  Roofline for the first 10 convolutional layers of VGG16 implemented with Winograd on RISC-VV@gem5}
  \label{fig:RooflineWinogradVGG16}
  \vspace{-1.3em}
\end{figure}

We perform a roofline model \cite{williams2009roofline} analysis for the convolutional layers of VGG-16 using our Winograd implementation, for the single core simulated by gem5 for RISC-VV with a 512-bit vector length and 1MB of cache, with peak GFLOPS as 64 GFLOPS/sec and memory bandwidth as 13GB/sec. We use the profiling data from gem5 to compute the arithmetic intensity (AI) for each convolutional layer of VGG-16, i.e. each layer includes all the transformation kernels along with the tuple multiplication kernel for all the layers of the VGG-16 network model. We note that we calculate the arithmetic intensity based on the DRAM bytes. For all the layers, we also validate our calculated GFLOPS against theoretically calculated GFLOPS for Winograd. Figure~\ref{fig:RooflineWinogradVGG16} shows the roofline model of the first 10 convolutional layers of the VGG-16 network model on gem5 for RISC-VV. Our roofline analysis shows that all the layers of VGG-16 are memory-bound, and this is the reason behind less scalability with longer vector lengths and large caches. However, our roofline analysis also shows that there is a scope for further improvement in the kernels of the Winograd algorithm as the values of the plots are far away from the memory bandwidth ceiling, therefore Winograd could benefit from cache-aware optimizations.  

\begin{figure}[!t]
  \centering
  \includegraphics[width=\linewidth]{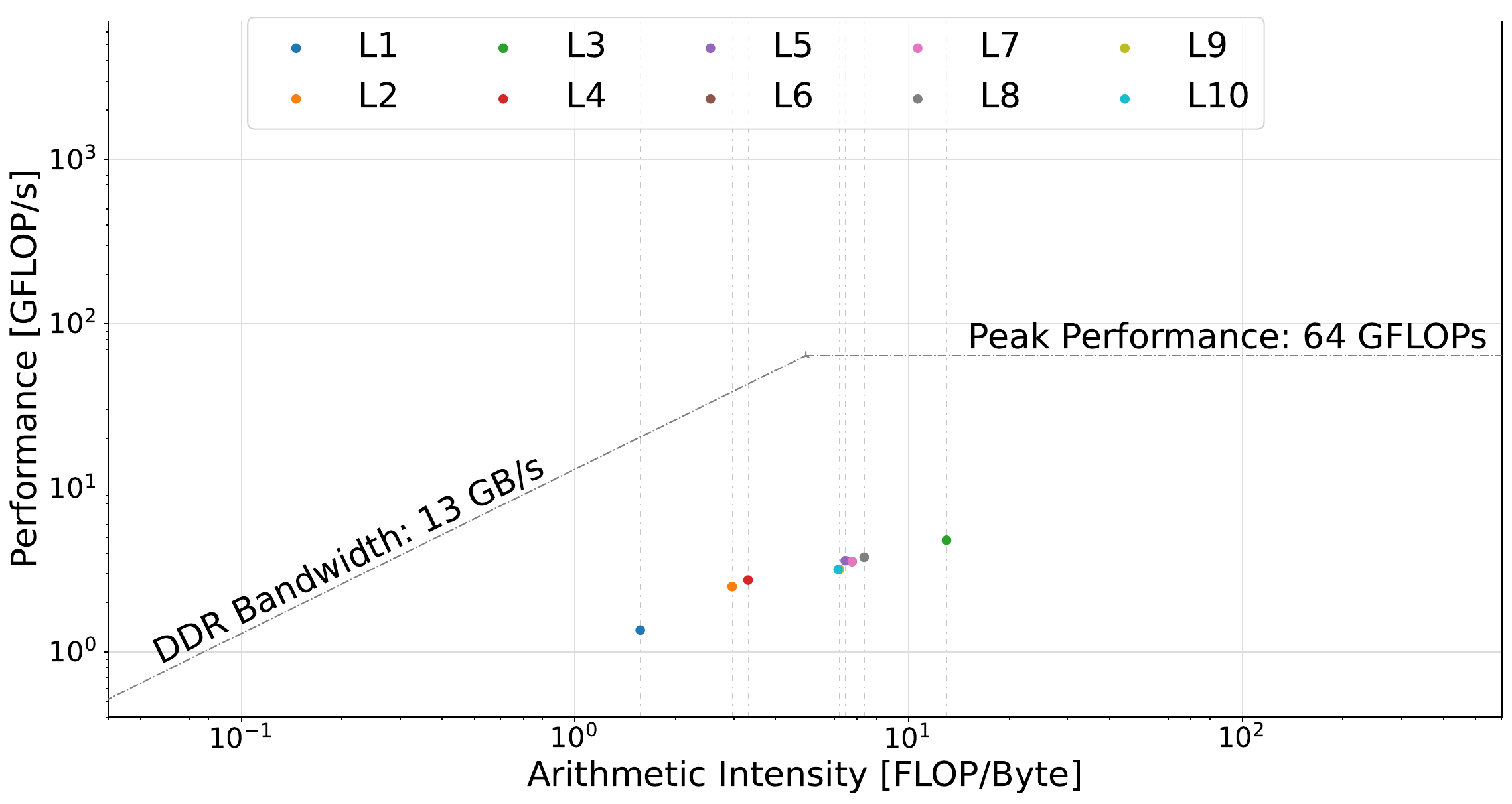}
  \caption{
     Roofline for the first 10 convolutional layers of VGG16 implemented with im2col+GEMM on RISC-VV@gem5}
  \label{fig:Rooflineim2colgemmVGG16}
  \vspace{-1.3em}
\end{figure}

Further, Figure~\ref{fig:Rooflineim2colgemmVGG16} shows a similar roofline analysis of the first 10 convolutional layers of VGG-16, implemented with im2col+GEMM, on gem5 for RISC-VV. This plot shows that only 3 layers out of 10 layers are memory-bound, while the rest of the layers are compute-bound. This also supports our co-design findings for YOLOv3, which scales better with longer vector lengths, as it uses our hybrid approach, where some of the convolutional layers are using im2col+GEMM. However, we note that, as in the case of Winograd, the achieved performance is far from the peak performance ceiling, therefore there is room for further optimization of this method on vector architectures.

\label{Roofline}

\section{Discussion on Tools}
Multiple emulators/simulators such as Spike~\cite{SPIKE}, Vehave~\cite{vehave}, and gem5~\cite{gem5benchmark} provide support for simulating the vector instructions for RISC-VV. Spike is a functional simulator normally used for validating the vectorized implementation and supports up to 4096-bit vector length. Vehave is an emulator used for validating the correctness of results and is useful for collecting execution traces. These traces can be used with the MUSA simulator~\cite{MUSASimulation} for RISC-VV for further performance evaluation and exploration. Vehave supports up to 16384-bit vector lengths for RISC-VV, however, it can be only used on RISC-V hosts. On the other hand, gem5 is a cycle-accurate simulator that provides detailed performance analysis and allows to tune micro-architectural parameters. Two gem5 forks for RISC-VV exist, one implementing the "V" extension v0.7~\cite{gem5benchmark} and one implementing the "V" extension v1.0~\cite{plct-gem5}. 

In our work, we opt to use the gem5 simulator, as it allows us to study the impact of several hardware parameters, such as vector lengths, cache sizes, and the number of vector lanes on performance. Additionally, unlike the Vehave/MUSA toolchain, which targets the RISC-V architecture, it does not require a specific host, and also allows us to perform the same analysis on different architectures (e.g. ARM-SVE). 
However, developing and maintaining the tools plays a vital role in evaluating the performance of the architecture and studying the crucial architecture parameters tuning for enhancing the performance. While working with RISC-VV we encountered limitations; the gem5 simulator for RISC-VV~\cite{gem5benchmark}, based on decoupled vector architecture, is no longer maintained and therefore does not support the latest version of the extension or latest compiler. The alternative fork of gem5~\cite{plct-gem5} for RISC-VV, used in this work, is based on an integrated vector architecture and very long vector architecture support is not available yet. Also, there is no coordination among tools for supported vector lengths. Such limitations make progress slow, both in software and hardware design.

We believe that the software ecosystem plays a vital role in the maturation of architectures and in assessing the porting efforts and the expected performance of ported and optimized kernels, especially when the architecture is not yet realized in real hardware. Moreover, the coordination of the evolution of the ISA with the compiler and emulation/simulation tools is important, to relieve the developer from the unnecessary burden of porting and maintaining low-level code, and to increase programmability, allowing the developer to benefit from the latest optimizations in compilers. 



\section{Conclusion}
In this paper, we presented our vectorized and optimized implementation of the Winograd algorithm, used to implement convolutional layers, on RISC-V, using the ``V" vector extension (RISC-VV) through intrinsic instructions. Our vectorization and optimization effort revealed that workarounds for contiguous loads/stores alongside slideup instructions are more performant than scatter/gather (indexed) loads/stores, and that strided vector instructions perform equally to scatter/gather instructions in the task of transposing a vector, as they both cannot avoid memory accesses. 
We evaluated our optimized Winograd for RISC-VV on gem5 on an inference task using the YOLOv3 and VGG16 network models, achieving $1.08\times$ and $1.76x$ speedup respectively, compared to using im2col+GEMM for the implementation of the convolutional layers. We performed a co-design study for the two network models, considering the vector length and L2 cache size as tunable hardware parameters. Our results showed that    
 Winograd's algorithmic implementation does not require very long vector lengths i.e. beyond 2048 bits, because Winograd reduces the computational complexity and the total required floating point operations for implementing the convolutional layer. Our analysis also showed that Winograd does not have a very large L2 cache requirement, and can scale effectively with up to 64MB of L2 cache.



From our experience with optimizing for RISC-VV, we emphasize the significance of ensuring coordination in the progress between tools, ISAs, and compilers, in order to facilitate porting and optimization, as well as design space exploration for future architectures. Without such coordination, it can prove challenging to navigate the software and hardware design space. 

\begin{acks}
This work has been supported by the Swedish Research
Council via registration number 2020-04892.  The simulations were enabled by resources provided by the National Academic Infrastructure for Supercomputing in Sweden (NAISS) at PDC partially funded by the Swedish Research Council through grant agreement no. \mbox{2022-06725}. \looseness=-1
\end{acks}

\bibliographystyle{ACM-Reference-Format}

\bibliography{biblio}
\end{document}